\documentclass[epj]{svjour}
\usepackage{graphicx}
\input{epsf} 
\usepackage{amssymb}

\begin{document}

\title{Chaotic delocalization of two interacting particles \\
in the classical Harper model}

\author{
D.L. Shepelyansky
}
\institute{
Laboratoire de Physique Th\'eorique du CNRS, IRSAMC, 
Universit\'e de Toulouse, CNRS, UPS, 31062 Toulouse, France
}

\titlerunning{Chaotic delocalization of two interacting particles in the 
classical Harper model}
\authorrunning{D.L.Shepelyansky}

\abstract{We study the problem of two interacting particles in the
classical Harper model in the regime when
one-particle motion is absolutely bounded 
inside one cell of periodic potential.
The interaction between particles breaks integrability
of classical motion leading to emergence of Hamiltonian
dynamical chaos. At moderate interactions 
and certain energies above the mobility edge
this chaos leads to a chaotic propulsion of two particles
with their diffusive spreading over the whole space
both in one and two dimensions.
At the same time the distance between particles remains
bounded by one or two periodic cells
demonstrating appearance of 
new composite quasi-particles called chaons.
The effect of chaotic delocalization
of chaons is shown to be rather
general being present for Coulomb and short range interactions. 
It is argued that such delocalized chaons can be
observed in experiments with cold atoms and ions
in optical lattices.
}


\date{Dated: April 12, 2016}

\maketitle

\section{Introduction}
\label{sec1}
The Harper model describes a quantum dynamics of an electron
in a two-dimensional periodic potential (2D) and a perpendicular 
magnetic field \cite{harper}. Due to 
periodicity of potential the problem can be reduced to the Schr\"odinger
equation on a discrete quasiperiodic one-dimensional (1D) lattice.
This equation is characterized by a dimensionless Planck constant $\hbar$
determined by a magnetic flux through the lattice cell.
The fractal spectral properties 
of this system have been discussed in  \cite{azbel}
and the fractal structure of its spectrum was 
directly demonstrated in \cite{hofstadter}.

For typical irrational flux values the system has
a Metal-Insulator Transition (MIT) 
established by Aubry and Andr\'e \cite{aubry}. The MIT takes place when
the amplitude $\lambda$ of the quasiperiodic potential (with hopping 
being unity) is changed from $\lambda <2$ (metallic or delocalized  phase) 
to $\lambda>2$ (insulator or localized phase).
A review of the properties of the Aubry-Andr\'e model can be found 
in \cite{sokoloff} and the mathematical prove of  MIT is given in 
\cite{lana1}. The stationary Schr\"odinger equation of the system
has the form
\begin{equation}
\begin{array}{c}
\lambda \cos(\hbar n+\beta)\phi_{n} + \phi_{n+1}+\phi_{n-1}=E\phi_{n}
\end{array}
\label{eq1}
\end{equation}
or in the operator representation
\begin{equation}
\begin{array}{c}
\hat{H}\psi = [\lambda \cos\hat{x}+ 2 \cos \hat{p}] \psi = E\psi \:\: ,
\end{array}
\label{eq2}
\end{equation}
where $\hat{p}, \hat{x}$ are momentum and coordinate operators
with the usual commutator $[\hat{p},\hat{x}]=-i\hbar$ \cite{sokoloff}.

\begin{figure}
\begin{center}
\includegraphics[width=0.24\textwidth]{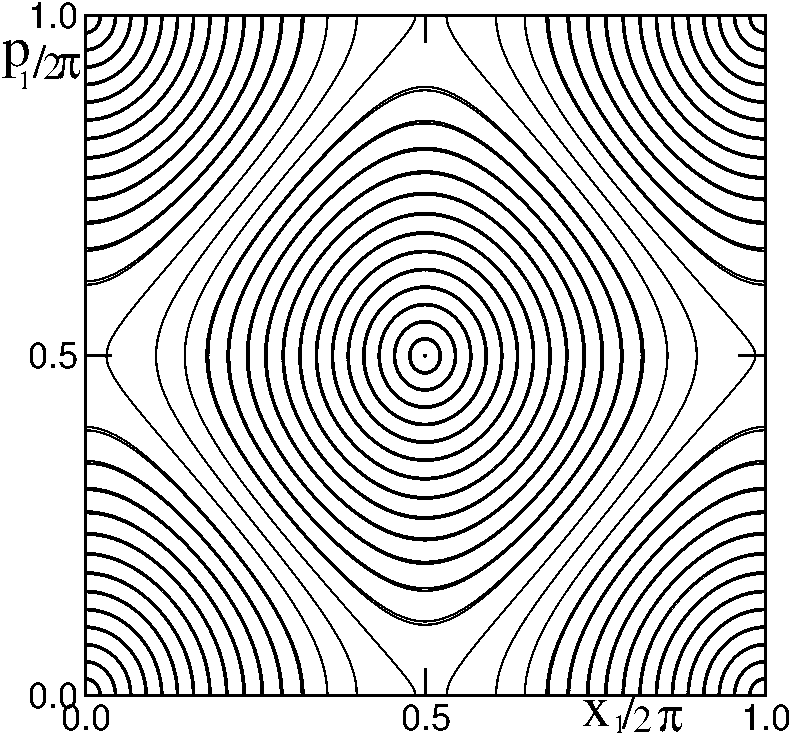}
\includegraphics[width=0.24\textwidth]{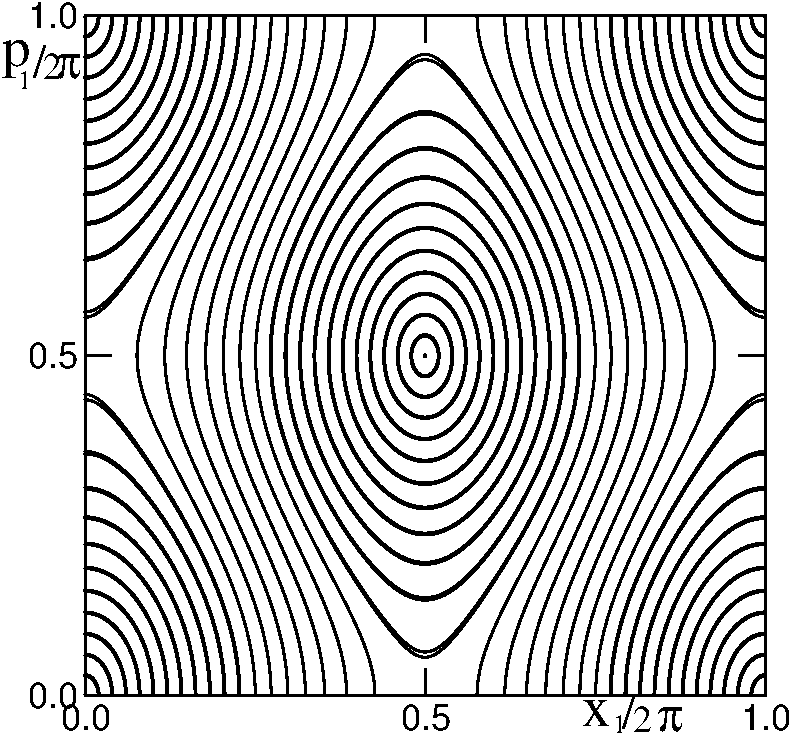}
\caption{Phase space of the one-particle classical Harper model
at $\lambda=2.5$ (left panel) and  $\lambda=4.5$ (right panel);
curves correspond to 60 trajectories with different initial conditions.
Only one cell of the periodic phase space is shown.
}
\label{fig1}
\end{center}
\end{figure}

From the view point of classical dynamics the critical
value $\lambda=2$ is very natural. Indeed, the dynamics is
described by the classical Hamiltonian 
\begin{equation}
\begin{array}{c}
H(p,x) = \lambda \cos {x}+ 2 \cos {p} = E \;\; ,
\end{array}
\label{eq3}
\end{equation}
with commuting conjugated variables $(p,x)$ in  (\ref{eq2}).
For $\lambda>2$
the maximal value of kinetic term $K=2\cos p$
is smaller than the potential barrier $V=\lambda \cos {x}$
and a particle cannot overcome the barrier
being localized in coordinate space
(all equipotential curves are vertical on the phase plane $(p,x)$).
In the opposite case  $\lambda<2$
the maximal value of potential barrier is smaller
than the kinetic term and at certain energies 
a particle can propagate
ballistically along $x$. 
Examples of  phase space curves for the classical localized phase
at $\lambda > 2$ are shown in Figure~\ref{fig1}.
The energy of the system is restricted to the interval
$-2-\lambda \leq E \leq 2+\lambda$. At $\lambda=2$ 
and $E=0$ the separatrix lines $p_1=x_1+\pi +2\pi m_1$,
$p_1=-x_1+\pi +2\pi m_2$ go to infinity
covering the whole phase space
($m_1, m_2$ are integers).  

Of course, the behavior of quantum system is much more
subtle due to presence of quantum tunneling
so that highly skillful mathematical methods are
required to prove quantum localization of 
eigenstates at typical irrational flux
value $\hbar/2\pi$ 
and to extend analysis to more general hopping terms
(see \cite{lana1,lana2,lana3}). The numerical studies of the
quantum model can be found at \cite{geisel,austin}.

The investigation of interaction effects between particles in the 
1D quantum Harper model was started in \cite{dlsharper} 
with the Hubbard interaction of 
Two Interacting Particles (TIP). 
It was found that the interaction  creates 
TIP localized states in the regime 
when all eigenstates of  noninteracting particles are delocalized
in the 1D Harper model
(metallic phase at $\lambda <2$). 
Further studies also found enhancement of localization effects
in presence of interactions \cite{barelli,orso}. 
This localization enhancement
is opposite to the TIP effect in disordered systems
where the interactions increase the TIP localization length in 1D 
\cite{dlstip,imry,pichard,frahm1995,vonoppen,frahm1999,frahm2016}
or even lead to delocalization of TIP pairs for dimensions $d \geq 2$ 
\cite{borgonovi,dlscoulomb,lagesring}.
Thus interactions between two particles  in systems with disorder
can even destroy the Anderson localization
existing for  noninteracting particles.
The tendency in the 1D Harper model seemed to be an opposite one.

Thus the results obtained in \cite{flach}
on the appearance of delocalized TIP pairs 
in the 1D quantum Harper model,
for certain particular values of interaction strength and energy, 
in the regime, when all one-particle states are exponentially localized,
is really surprising. The recent advanced studies
confirmed the existence of 
so called Freed by Interaction Kinetic States (FIKS)
with delocalized quasiballistic FIKS pairs,
existing at  various irrational flux values,
propagating over the whole large system sizes  \cite{fiks1d}.
The studies of TIP on the 2D Harper model
showed the presence of subdiffusive delocalization of TIP
but found no signs of quasiballistic states \cite{fiks2d}.

At present the skillful experiments
with cold atoms in optical lattices allowed 
to realize the 1D Harper model,
to observe there the MIT transition
for noninteracting particles and to
perform studies of interactions \cite{roati,modugno,bloch}.
The first steps in experimental study
of the 2D Harper model are reported
recently \cite{bloch2d}. 

The Harper Hamiltonian (\ref{eq2})
appears also in such solid-state systems like
incommensurate crystals where
a free electron propagation in 
a finite energy band ( $E=2\cos p$)
is affected by atomic charges 
creating an effective periodic potential
( $V(x)= \lambda \cos x$) \cite{pokrovsky,gruner,brazovski}.
Indeed, the energy band spectrum 
like $E \sim \cos p$ naturally appears
in semiconductor heterostructures and superlattices
(see e.g. \cite{alper}). Hence, the investigation of
interaction effects in the Harper model
can be relevant also for incommensurate 
crystals.

In view of this theoretical and experimental progress it
is important to obtain a better understanding of the
physical origins of FIKS pairs and TIP delocalization
in the Harper model. With this aim we study the
properties of TIP in the classical Harper model
considering the Hamiltonian dynamics in the 
classical conservative systems with two (and four) degrees of freedom
in the 1D (and 2D) Harper model. We show that 
at rather generic conditions
the interactions destroy classical integrability of 
motion and localization, 
leading to chaos and chaotic propulsion
of TIP characterized by a diffusive spreading in
coordinate space.

The paper is composed as follows:
Section 2 describes TIP with Coulomb interactions
in the 1D Harper model,
Section 3 describes TIP with a short range interaction
in 1D, Section 4 describes TIP with Coulomb interactions
in the 2D Harper model and the discussion of the results is
presented in Section 5.

\section{TIP with Coulomb interactions in 1D}
\label{sec2}

The classical TIP Hamiltonian in the 1D Harper model reads:
\begin{equation}
\begin{array}{c}
H(p_1,p_2,x_1,x_2) = 2(\cos p_1 +\cos p_2)  \\
          + \lambda (\cos x_1 + \cos x_2) 
          + U/((x_2-x_1)^2+b^2)^{1/2} \;\; .
\end{array}
\label{eq4}
\end{equation}
Here $U$ is a strength of Coulomb interaction
and $b$ is a certain screening or regularization length
appearing due to quantum smoothing or  
effective finite distance in 2D. In the following
we keep $b=1$ since the results are not very sensitive
to $b$ value as soon as $b$ is smaller than the 
lattice period $d=2\pi$ in space.
Here we use dimensionless units
where the lattice spatial  period is $d=2\pi$.
In physical units the interaction
strength can be measured as
$U=2\pi e^2/d$ where $e$ is electron charge and $d$ 
the lattice period.  

\begin{figure}
\begin{center}
\includegraphics[width=0.23\textwidth]{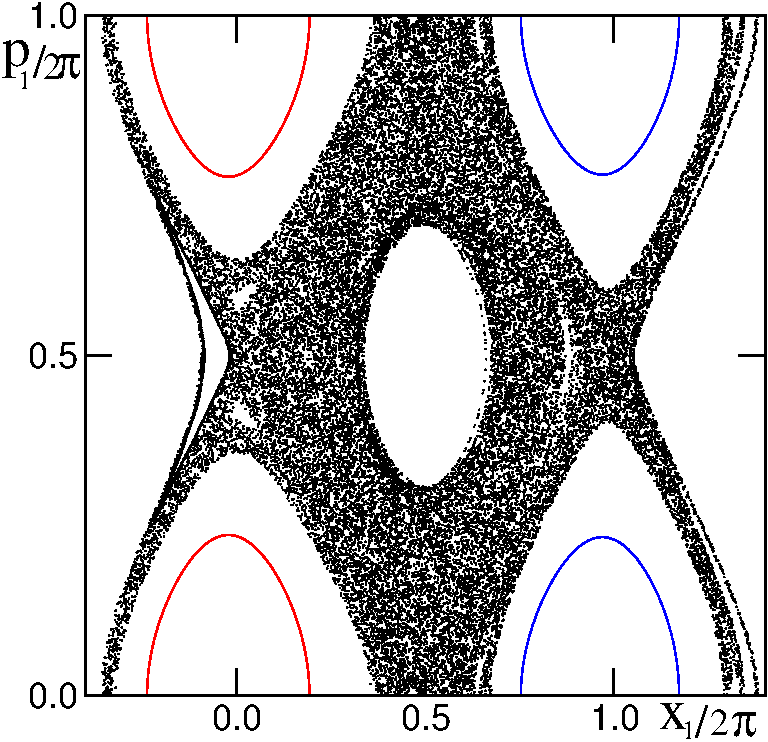}
\includegraphics[width=0.24\textwidth]{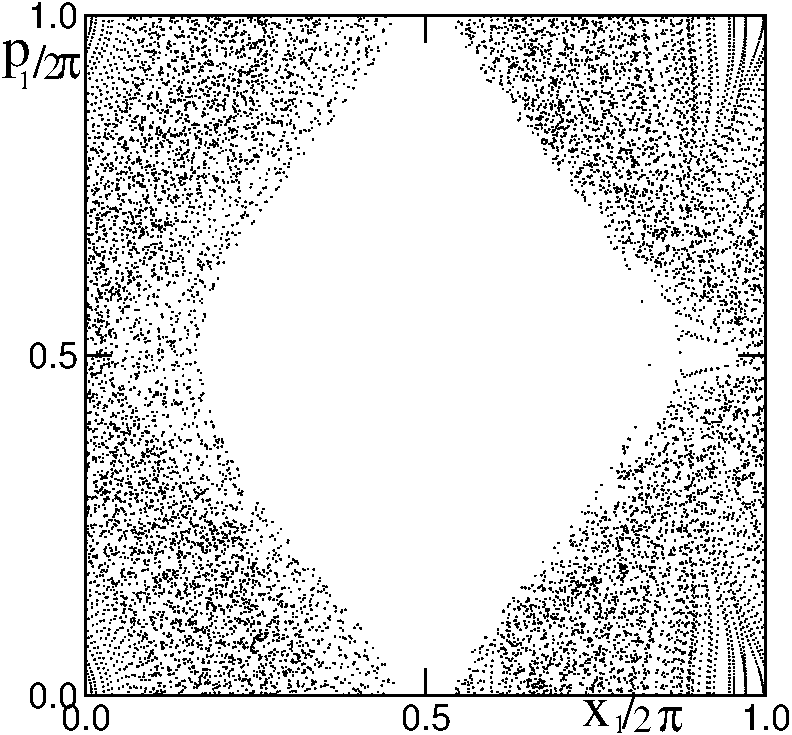}
\caption{Poincar\'e sections for TIP at $\lambda=2.5$ and $U=1$;
left panel: 3 trajectories are shown
at energies $E=-2.4$ (black points),
$-0.046$ (red points), $-0.07$ (blue points)
up to times $t=2 \times 10^5$;
right panel: one trajectory at
$E= 0.42$ (black points) up to $t=10^5$.
The vertical  axis shows the fractional part
of $p_1/2\pi (mod \; 1)$ and 
the horizontal axis shows $x_1/2\pi$  
(left panel, no fraction) and the fractional part of 
$x_1/2\pi (mod  \; 1)$ (right panel,
$x_1/2\pi$ varies in the range $(-12,1)$);
the sections are taken at a fractional part $p_2/2\pi=1/2$
and $dp_2/dt>0$; white area in the right panel corresponds 
to energy forbidden region.
}
\label{fig2}
\end{center}
\end{figure}

At $U=0$ we have integrable dynamics of noninteracting particles
which is bounded in space $x$ inside one periodic cell
at any energy. For finite $U$ values 
the invariant Kolmogorov-Arnold-Moser (KAM)
curves start to be destroyed by interactions with appearing of
chaotic motion \cite{chirikov,lichtenberg} 
with unbounded diffusion of pairs in $x$ space. 
It is clear that particles can diffuse only in pairs
since individual particles are localized inside
one lattice period due to integrability of one-particle dynamics
and energy restrictions discussed above.
From the energetic viewpoint a kinetic energy of
two particles $K= 2 (\cos p_1 +\cos p_2)$
can become larger than a potential barrier
$V=\lambda(\cos x_1 +\cos x_2)$
in effective 2D space of TIP
that can allow to overcome this barrier
leading to extended propagation of TIP.
We call such delocalized TIP pairs chaons
since they are generated by chaos.

We note that in presence of interactions the allowed
energy band of the Hamiltonian (\ref{eq4})
is $-4-2\lambda \leq E \leq 4+2\lambda +U$
(due to band energy structure
and symmetry $p \rightarrow -p, x \rightarrow -x$
we consider only the repulsive case $U \geq 0$; 
the attractive case $U<0$ has the same 
behavior as at $U>0$).

\begin{figure}
\begin{center}
\includegraphics[width=0.24\textwidth]{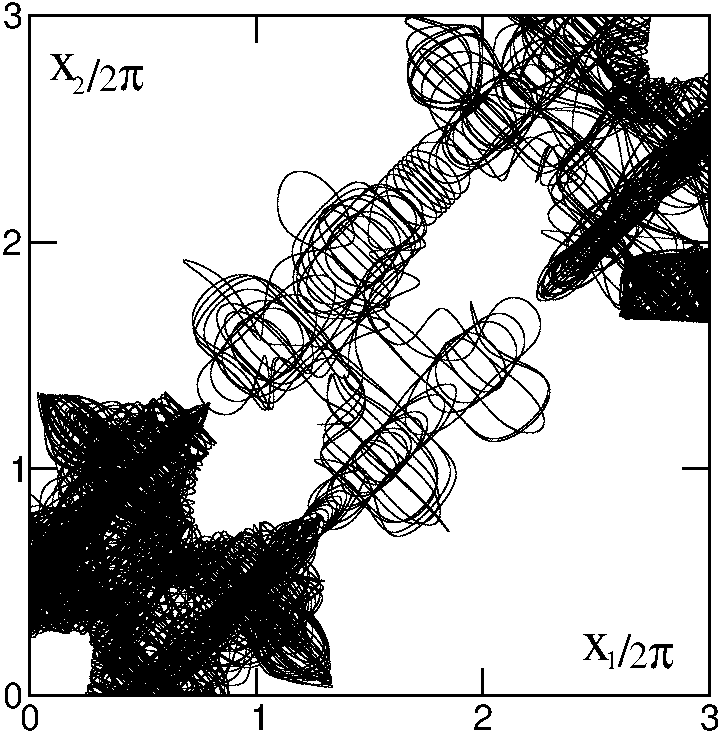}
\includegraphics[width=0.24\textwidth]{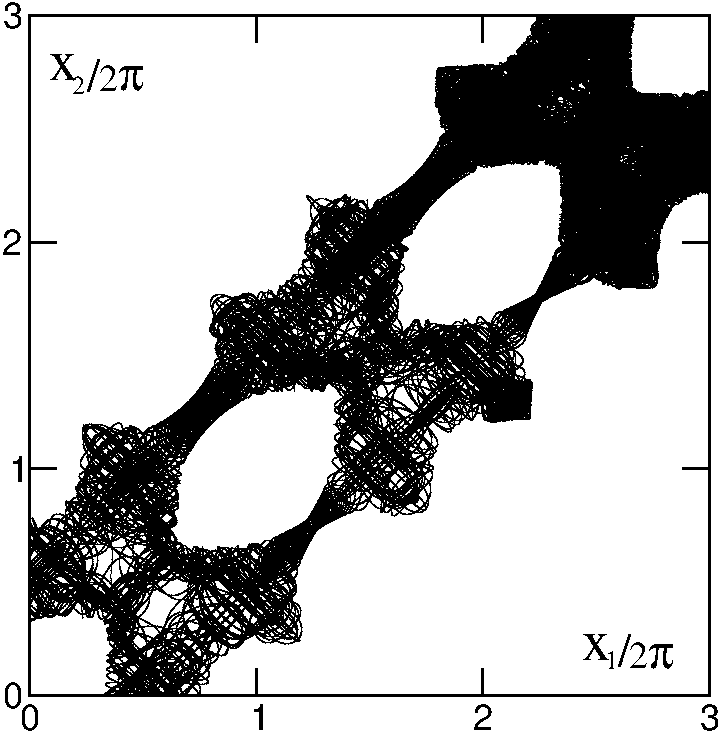}\\
\includegraphics[width=0.24\textwidth]{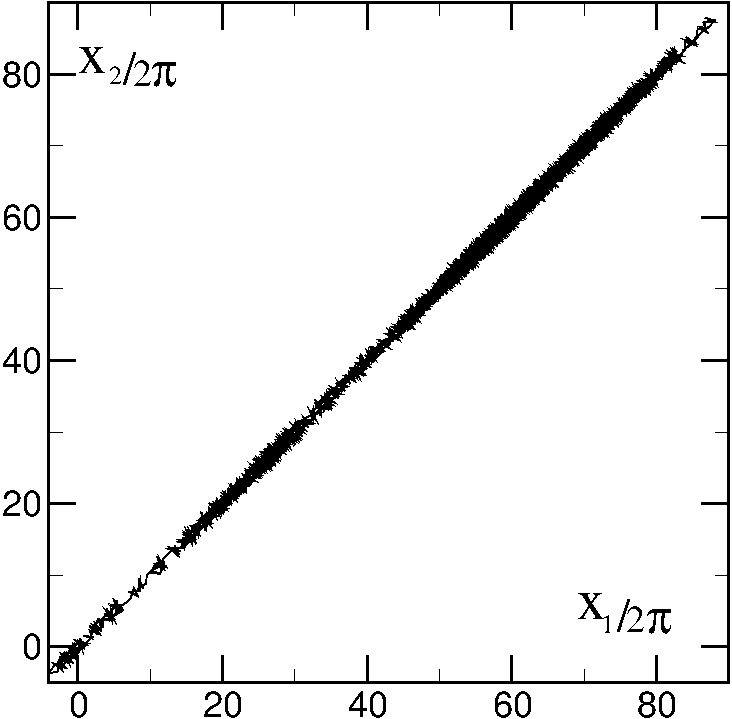}
\includegraphics[width=0.24\textwidth]{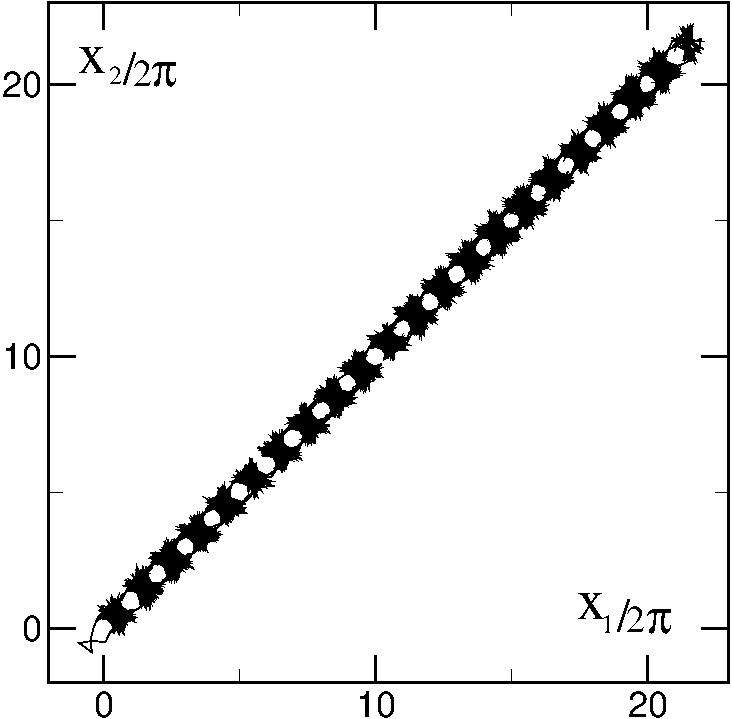}\\
\includegraphics[width=0.24\textwidth]{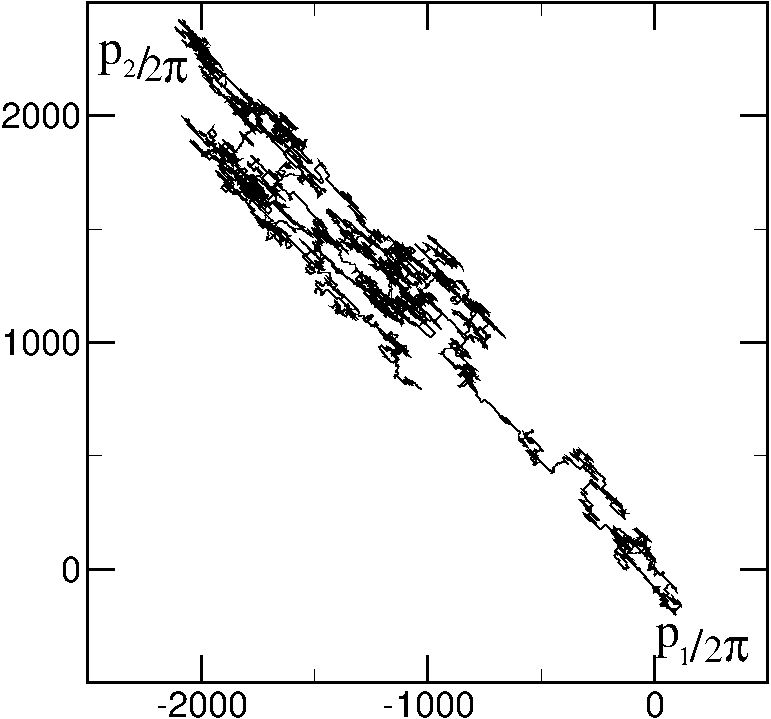}
\includegraphics[width=0.24\textwidth]{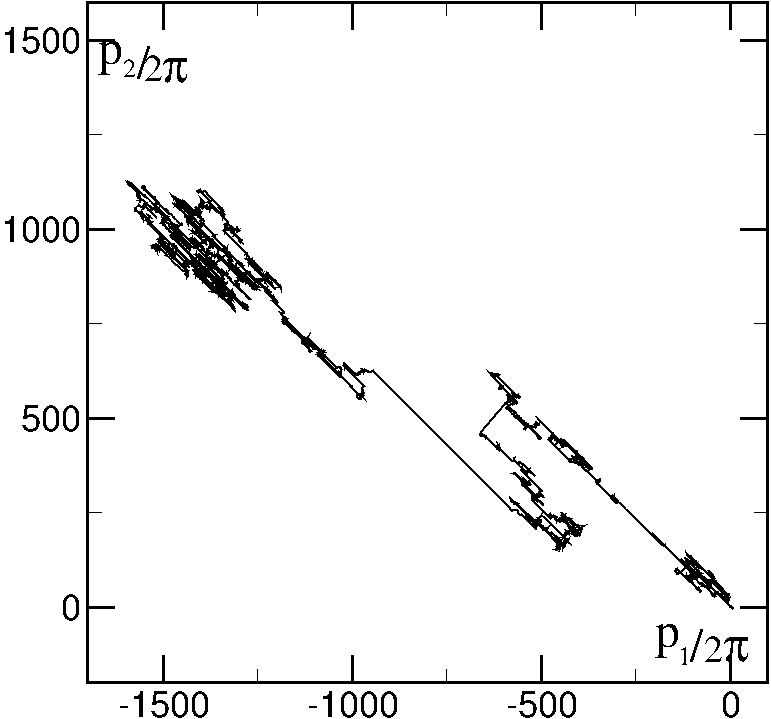}
\caption{Dynamics of a typical TIP trajectory in the classical Harper model
is shown for $\lambda=2.5$, $U=6$, $E=1.884$ (left column) and
$\lambda=4.5$, $U=12$, $E=3.876$ (right column).
Top row shows TIP dynamics in plane $(x_1, x_2)$ on small scale;
middle row shows  the same on large scale up to times
$t=10^6$; bottom row shows TIP dynamics in $(p_1,p_2)$ 
on large scale up to times $t=10^6$;
positions are shown with a time step $0.01$
(top) and $10$ (middle, bottom).
}
\label{fig3}
\end{center}
\end{figure}

The Hamiltonian dynamics of (\ref{eq4})
is integrated numerically by the  Runge-Kutta method
with a typical time step $\Delta t =0.01$
and the relative accuracy of energy
conservation being around $10^{-8}$
for times $t \sim 100$ and better than $10^{-5}$
for $t \sim 10^{6}$.

Examples of Poincar\'e sections \cite{lichtenberg}
at a moderate interaction $U=1$ are shown in Figure~\ref{fig2}
for $\lambda=2.5$ when noninteracting particles are
localized inside one coordinate cell.
At some energies the dynamics remains integrable
($E=-0.046, -0.07$), it can be also chaotic
but bounded inside one or two coordinate cells
($E=-2.4$) or to be chaotic and unbounded
in coordinate space ($E=0.42$).
The emergence of chaos induced by interactions is
rather natural since the one-particle system is 
strongly nonlinear.

Typical examples of trajectories in coordinate and momentum space
are shown in Figure~\ref{fig3} for $\lambda=2.5$ and $4.5$.
On a scale of a few cells there is a complex,
chaotic dynamics of TIP inside a given cell
leading to chaotic transitions to nearby cells (top row).
In this manner a chaotic propulsion of TIP
generates TIP propagation along $x-$axis.
In all cases of delocalized TIP at $U \leq 20$ the distance
between particles is not exceeding
$\Delta x_M= \max|x_2-x_1| =2$ (middle row).
While in $x-$direction the propagation is diffusive
(see Figures~\ref{fig4},~\ref{fig5} below)
the spreading in momentum remains 
quasiballistic with approximately linear growth
of $p_1, p_2$ with time (bottom row);
for momentum this is not very surprising since
there is a ballistic growth of $p$
even for one particle (see Figure~\ref{fig1}).

\begin{figure}
\begin{center}
\includegraphics[width=0.24\textwidth]{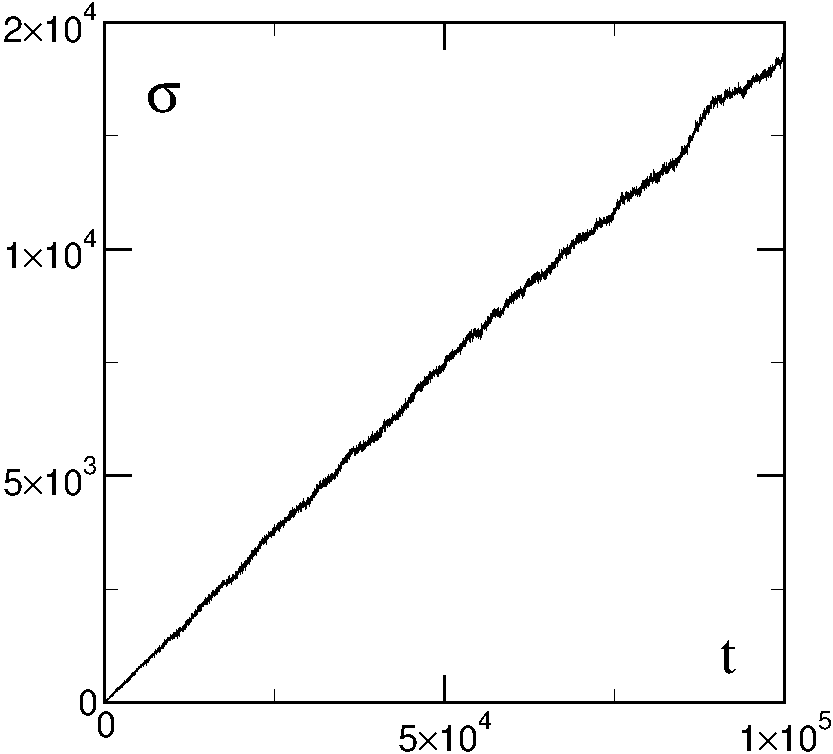}
\includegraphics[width=0.24\textwidth]{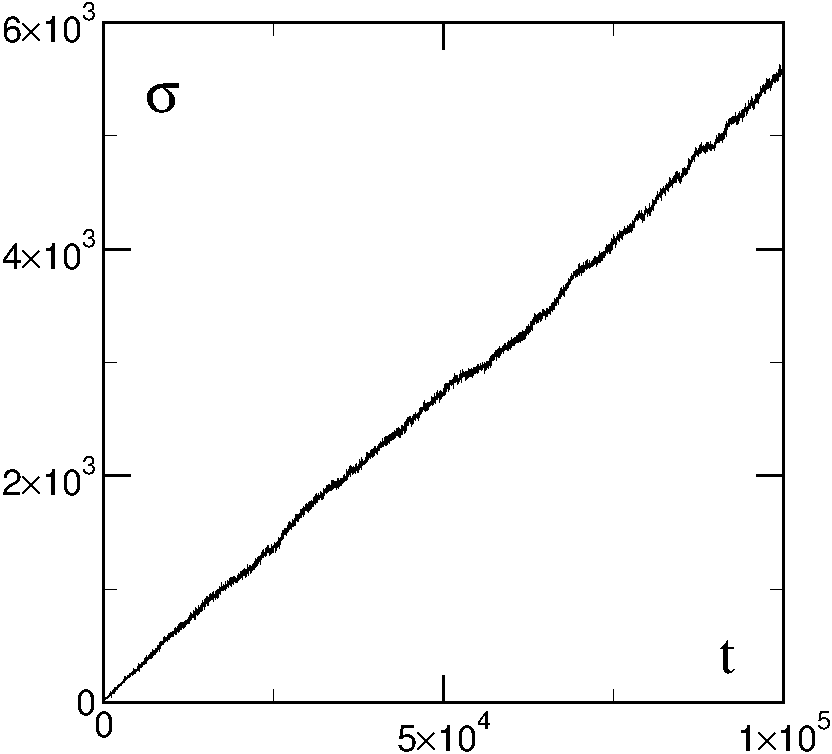}
\caption{Time dependence of the second moment
$\sigma(t)= <((x_1)^2+(x_2)^2)/2 >$
averaged over $10^3$ orbits taken
in a vicinity of trajectories of Fig.~\ref{fig2}
at $\lambda=2.5$, $U=6$, $E \approx 1.884$ (left panel)
and  $\lambda=4.5$, $U=12$, $E \approx 3.876$ (right panel).
The fit gives the diffusion rate $\sigma = Dt + const$
with $D=0.143 \pm 3.2 \times 10^{-5}$  (left panel) and
$0.0543 \pm 1.4 \times 10^{-4}$ (right panel).
}
\label{fig4}
\end{center}
\end{figure}

The diffusive nature of TIP propagation and
spreading along $x$ is directly illustrated in
Figures~\ref{fig4},~\ref{fig5}. Indeed, the second moment
$\sigma=<({x_1}^2+{x_2}^2)/2>$ grows linearly with time
as $\sigma=Dt$, both for
$\lambda=2.5$ and $4.5$ (see Figure~\ref{fig4}).
These data are averaged over $10^3$ trajectories.

\begin{figure}
\begin{center}
\includegraphics[width=0.24\textwidth]{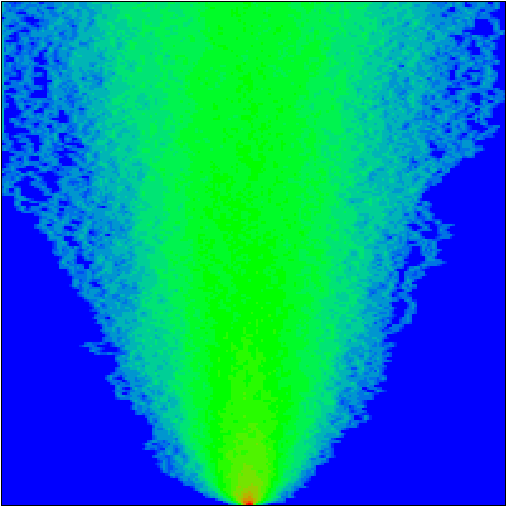}
\includegraphics[width=0.24\textwidth]{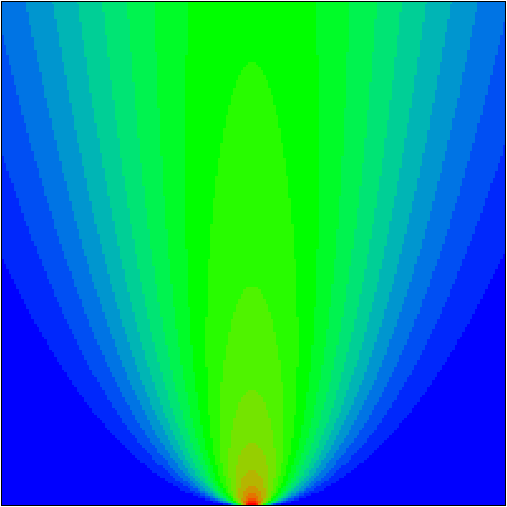}
\caption{One-particle density distribution
over coordinate range
$(-60 \leq x_{1,2}/2\pi \leq 60)$
on horizontal axis
and time interval $0 \leq t \leq 10^5$
on vertical axis.
Left panel: data are averaged over $10^3$ TIP orbits of Fig.~\ref{fig3}
at $\lambda=2.5$, $U=6$, $E \approx 1.884$;
right panel: the theoretical distribution 
$W(x,t) = \exp(-x^2/2Dt) / \sqrt{2 \pi Dt}$ of
the diffusion equation (\ref{eq5}) with the diffusion rate
$D=0.143$ of Fig.~\ref{fig4} 
shown in the same range as in left panel;
color is proportional to density
changing from zero (blue) to maximum (red).
}
\label{fig5}
\end{center}
\end{figure}

The probability or density distribution 
W(x,t) of these trajectories in $x$
is well described by the Fokker-Planck equation
\begin{equation}
\partial W(x,t)/ \partial t = (D/2)  \partial^2 W(x,t)/\partial x^2 \;\; .
\label{eq5}
\end{equation}
Indeed, the analytic solution is in a good agreement 
with the numerical result obtained by averaging over
$10^3$ trajectories of (\ref{eq4})
at times $t \leq 10^5$ (see Figure~\ref{fig5}).

\begin{figure}
\begin{center}
\includegraphics[width=0.24\textwidth]{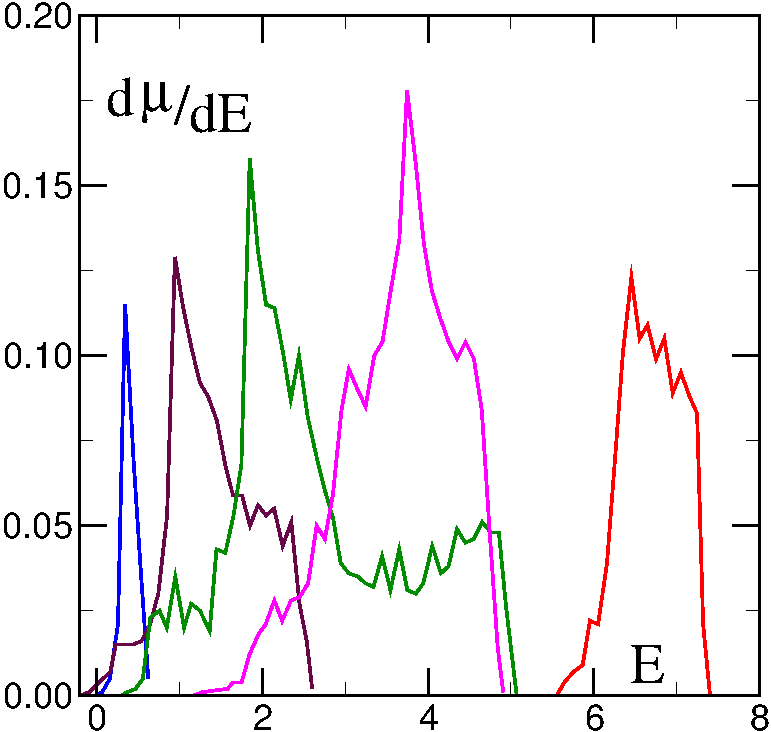}
\includegraphics[width=0.24\textwidth]{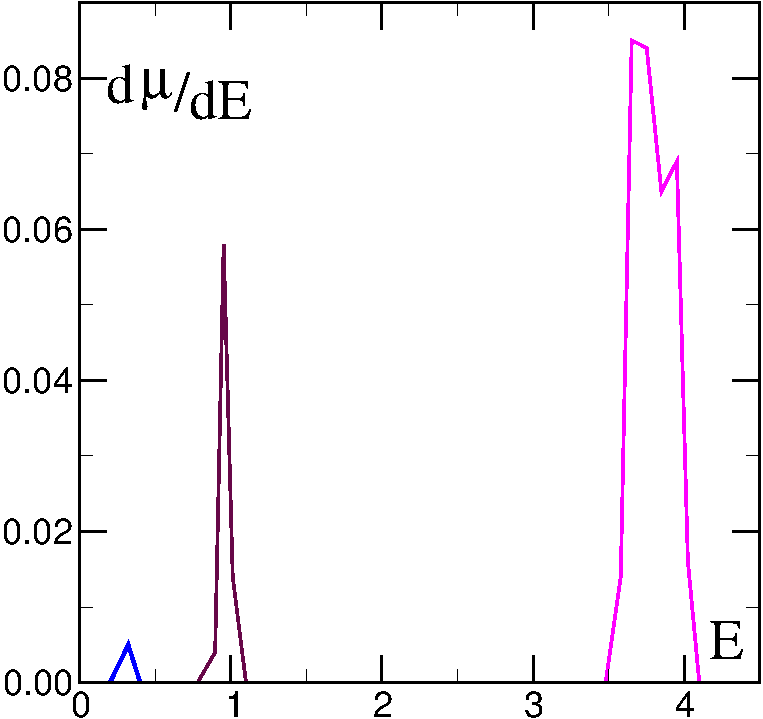}
\caption{Dependence of differential measure
of diffusive trajectories $d \mu/d E$ on TIP energy $E$
for: $\lambda=2.5$ and $U=1$ (blue), $3$ (brown), $6$ (green),
$12$ (magenta),  $20$ (red) with curves from left to right
with corresponding total measure 
$\mu =0.0252, 0.1319, 0.2295, 0.2420, 0.1190$ (left panel);  
$\lambda=4.5$ and $U=1$ (blue), $3$ (brown), $12$ (magenta)
with corresponding total measure $\mu = 0.0005, 0.0076, 0.0333$
(right panel). Data are obtained from $N=10^4$ trajectories
homogeneously distributed in the phase space and
iterated till time $t=10^6$; averaging is done over energy interval 
$\Delta E=0.1$. 
}
\label{fig6}
\end{center}
\end{figure}

\begin{figure}
\begin{center}
\includegraphics[width=0.24\textwidth]{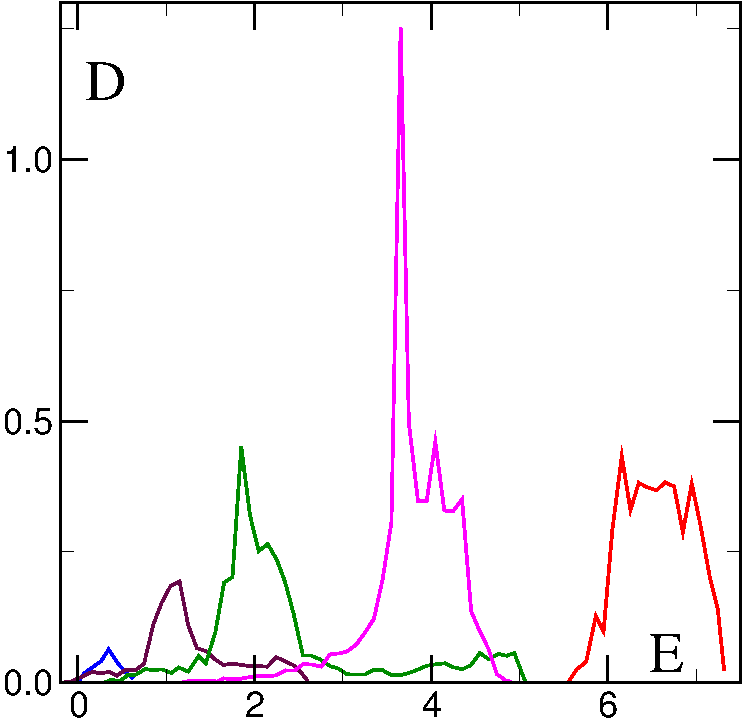}
\includegraphics[width=0.24\textwidth]{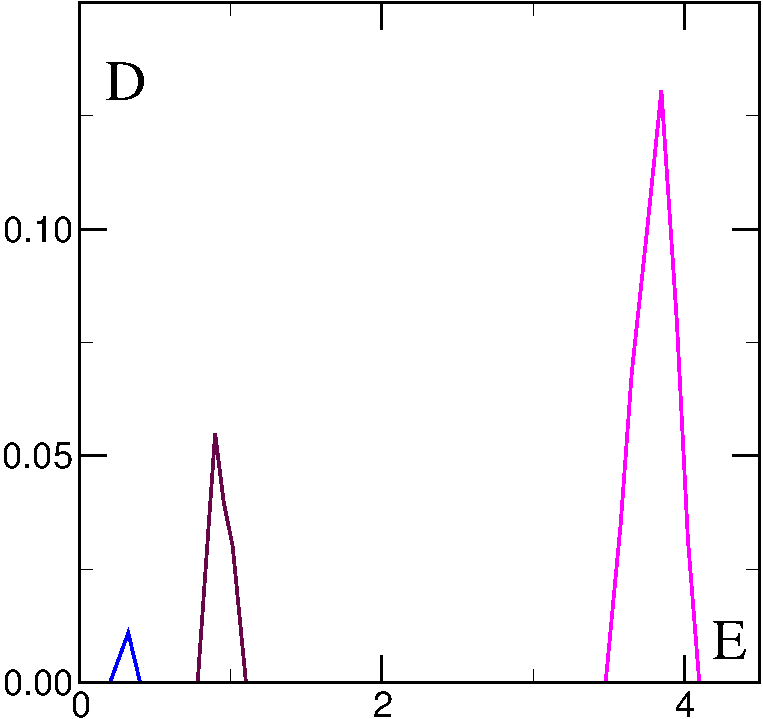}
\caption{Dependence of diffusion rate $D$ on TIP energy $E$
for same trajectories as in Fig.~\ref{fig6}
at same parameters and  colors with
$\lambda=2.5$ (left panel),
$\lambda=4.5$ (right panel).
}
\label{fig7}
\end{center}
\end{figure}

To determine the dependence of measure $\mu$ and diffusion coefficient $D$
on TIP energy $E$ we follow $N=10^4$ trajectories
till time $t=10^6$.
Initially at $t=0$ all trajectories are  homogeneously distributed
in the phase space with TIP being inside the same cell.
 Those trajectories with displacements
from initial positions being less then the size of 2 cells 
are considered as non-diffusive. The total measure $\mu$ of 
diffusive trajectories at all energies $N_d$ is determined 
as a fraction $\mu=N_d/N$. The differential distribution
$d \mu / d E$ is obtained from a histogram with energy interval
$\Delta E =0.1$. The obtained dependence $d \mu/d E$ on $E$ 
is shown in Figure~\ref{fig6} for
$\lambda=2.5; 4.5$ at different interactions.
The total measure $\mu$ is rather small at weak interactions
($\mu=0.025$ at $U=1, \lambda=2.5$), it increases till 
optimal values of $U \approx 12$ being comparable 
with the total energy band width $E_B= 8+4\lambda=18$,
and then decreases with $U$. The maximum of
$d \mu/d E$ is located approximately at energy
$E_{max} \approx U/\pi$ since at $U=0$ the most deformed KAM
curves are located at $E \approx 0$ and on average the distance
between particles inside one cell is $\pi$
giving a corresponding energy shift.

We note that for $\lambda=2.5, U=6$ the measure of diffusive orbits 
decreases from $\mu = 0.2295$ down to $\mu =0.09$ if the initial positions
of second particle are taken in nearby cell ($x_2 \rightarrow x_2+2\pi$).
Indeed, on average this gives a reduction of interactions
decreasing the measure of chaos. We should note that
at these parameters
for all diffusive trajectories the maximal
separation of particles during time evolution is
not exceeding the size of two cells 
($|x_2-x_1|/2\pi <2$).

The energy dependence of diffusion coefficient
$D(E)$ is shown in Figure~\ref{fig7}.
The maxima of $D$ are approximately at the same energies
as in Figure~\ref{fig6},
corresponding to developed chaos leading to TIP transitions
between nearby cells. Since only a fraction of trajectories
are delocalized the fluctuations of $D$ from one trajectory to another
are significant but averaging over
trajectories inside histogram interval
gives a reduction of such fluctuations.
The diffusion rate can be also estimated as
$D \approx (2\pi)^2/t_c$ where $t_c$ is an average
transition time between two cells.
With typical value $D \sim 0.25$
we obtain $t_c/2\pi \sim 25$ being significantly larger
then the period of small oscillations of one particle
in a vicinity of potential minimum in (\ref{eq2}).
This shows that many TIP collisions are required
to allow a jump from one cell to another.

For $\lambda=4.5$  the results of Figures~\ref{fig6},~\ref{fig7}
show that the measure of diffusive trajectories $\mu$ 
and their diffusion coefficient $D$ are strongly reduced.
Indeed, in this case the sum of kinetic terms of two particles is smaller
than the potential barrier of one particle and
thus the transitions of TIP from one cell to another
can take place only in narrow regions of phase space
(see Figure~\ref{fig3}). This leads to small values of $\mu$ and $D$.

In view of a strong growth of $p_1, p_2$
and their large separation growing with time
(see Figure~\ref{fig3} bottom panels)
it is clear that, for the original Harper system
of  charged particles in 2D potential (2$\cos y + \lambda \cos x$)
and a perpendicular magnetic field, there is no formation
of diffusive TIP pairs due to separation
of particles in $y$ direction which is 
analogous to $p$ in (\ref{eq3}). 

\section{TIP with short range interactions in 1D}
\label{sec3}

The one particle Hamiltonian (\ref{eq3}) is strongly nonlinear
and it is clear that practically any kind of interaction
$U_2(x_1,x_2)$ between particles should leads to
appearance of chaotic propulsion of TIP in coordinate space
at $\lambda >2$  when all one-particle orbits are bounded to one cell. 

As an example we consider the short range interaction
\begin{equation}
\begin{array}{c}
U_2(x_1,x_2) = U \cos^2(\pi (x_2-x_1)/2b) \;, \; |x_2-x_1| \leq b \;; \\
U_2=0 \; ,  \; |x_2-x_1| > b \; .
\end{array}
\label{eq6}
\end{equation}
As in the previous section we use $b=1$.

\begin{figure}
\begin{center}
\includegraphics[width=0.24\textwidth]{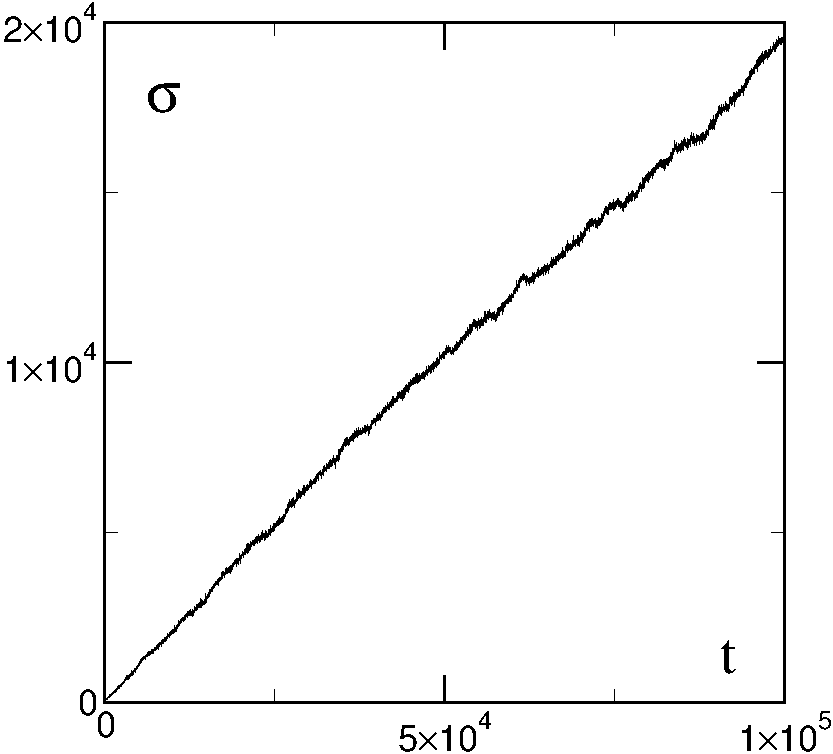}
\includegraphics[width=0.22\textwidth]{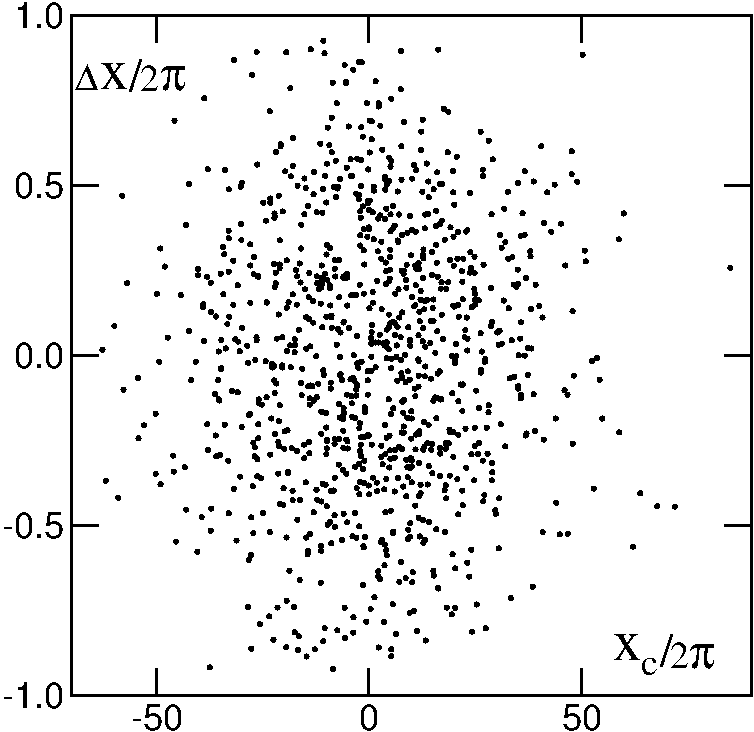}
\caption{Left panel: time dependence 
of the second moment
$\sigma(t)= <((x_1)^2+(x_2)^2)/2 >$
for the short range interaction model (\ref{eq6});
data are averaged over $10^3$ orbits taken
in a chaotic component
in a vicinity of energy $E=3.32$ 
at $\lambda=2.5$, $U=6$,
the fit gives the diffusion growth $\sigma = Dt + const$
with $D=0.189 \pm 6 \times 10^{-5}$.
Right panel: the distribution of $\Delta x = (x_2-x_1)$
and $x_c= (x_1+x_2)/2$ is shown for orbits of
left panel at $t=10^5$.
}
\label{fig8}
\end{center}
\end{figure}

For this model an example of diffusive growth of the second moment
$\sigma$  is shown in Figure~\ref{fig8} for
$\lambda=2.5$, $U=6$ and $E=3.32$. For these values of $\lambda$ and $ U$ 
the measure of delocalized orbits is found to be 
approximately $\mu \approx 0.1$
with $D \approx 0.15$ in the energy range
$1 < E< 5$ and close to zero outside.
At these parameters the separation between particles
does not exceed the cell size (see Figure~\ref{fig8}, right panel).

These results demonstrate that
the chaotic delocalization of TIP 
in the 1D Harper model appears also 
in the case of short range interactions.

\section{TIP with Coulomb interactions in 2D}
\label{sec4}

Let us now consider the dynamics of TIP
in the 2D classical Harper model.
Without interactions the Hamiltonian is the 
sum of the Hamiltonian (\ref{eq3})
in $x$ and $y$ directions. 
As in 1D the dynamics of each particle is bounded to a
one periodic cell.
In presence of smoothed Coulomb
interaction the TIP Hamiltonian has the form:
\begin{eqnarray}
&H&(p_{x_1},p_{x_2},p_{y_1},p_{y_2},x_1,x_2,y_1,y_2) = \\
\nonumber
&2&(\cos p_{x_1} +\cos p_{x_2} + \cos p_{y_1} +\cos p_{y_2})  \\
\nonumber
&+& \lambda (\cos x_1 + \cos x_2 + \cos y_1 + \cos y_2) \\
\nonumber
          &+& U/((x_2-x_1)^2 + (y_2-y_1)^2+b^2)^{1/2} \;\; .
\label{eq7}
\end{eqnarray}
As for 1D case we taken $b=1$ in the following
and $U \geq 0$.
The available energy band is restricted to the interval
$-8-4\lambda \leq E \leq 8+4 \lambda+U$.

\begin{figure}
\begin{center}
\includegraphics[width=0.24\textwidth]{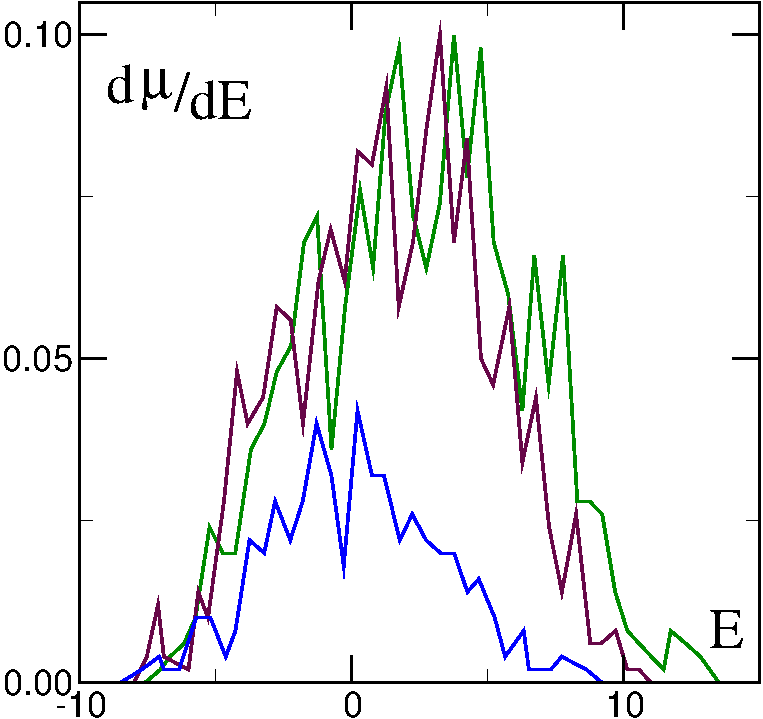}
\includegraphics[width=0.24\textwidth]{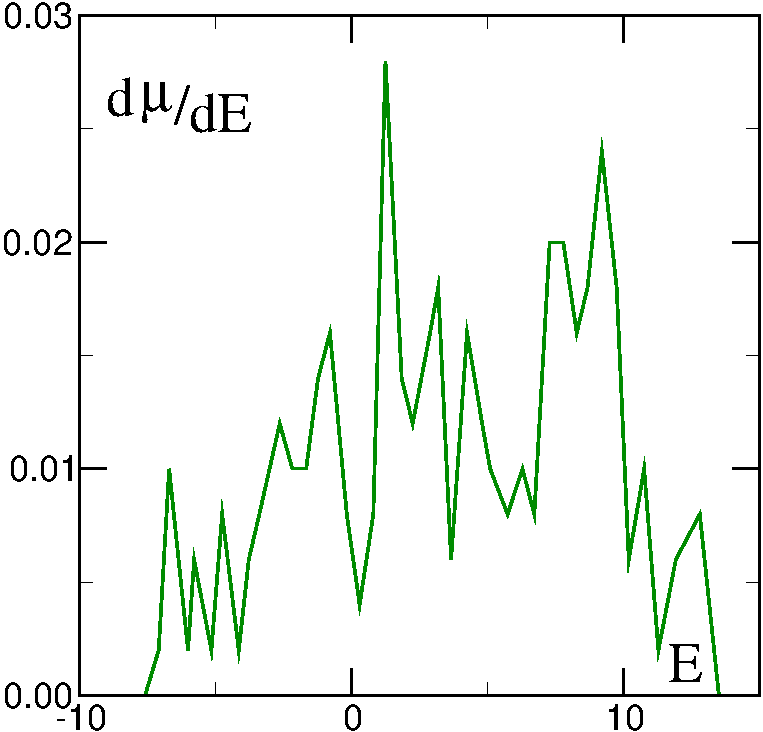}
\caption{Dependence of differential measure
of diffusive trajectories $d \mu/d E$ on TIP energy $E$ 
in the 2D Harper model
for: $\lambda=2.5$ and $U=1$ (blue), $3$ (brown), $6$ (green)
with corresponding total measure 
$\mu =0.265, 0.796, 0.893$ (left panel);  
$\lambda=4.5$ and $U=6$ (green)
with corresponding total measure $\mu = 0.217$,
for $U=3; 1$ the measure is small being respectively
 $\mu=0.023; 0$ and these date are not shown
(right panel). Data are obtained from $N=10^3$ trajectories
homogeneously distributed in the phase space and
iterated till time $t=4 \times 10^5$; averaging is done over energy interval 
$\Delta E=0.5$. 
}
\label{fig9}
\end{center}
\end{figure}

As for 1D the dynamics is integrated numerically with 
approximately the same accuracy. The measure of diffusive
trajectories $\mu$ is obtained from $1000$ orbits 
homogeneously distribution inside one cell
followed till times $t=4 \times 10^5$.
The diffusive trajectories are defined as those
that have a displacement larger than 3 cells
during this time. The dependence of $d \mu/dE$ is shown in
Figure~\ref{fig8} for $\lambda=2.5; 4.5$ at
$U=1, 3, 6$. The total measure at $U=6, \lambda=2.5$  
is increased comparing to the 1D case from
$\mu =0.23$ (1D, Figure~\ref{fig6}) to
$\mu =0.89$ (2D, Figure~\ref{fig9}).
Indeed, due to a larger number of degrees of freedom 
in 2D the measure of chaos increases.
However, for  $\lambda=4.5$ it becomes more difficult 
to penetrate through high potential barrier and there are
practically no diffusive TIP at 
$U=1, 3$. 

The data of Figure~\ref{fig9}
show the existence of an approximate mobility edge
with diffusion inside the energy
interval $E_{c_1} \approx -8 < E < E_{c_2} \approx 12$
(e.g. for $\lambda=2.5, U=6$). 
Of course, we should note
that in 2D there are 4 degrees of freedom and 
the Arnold diffusion can still lead
to a small measure of diffusive orbits
along chaotic separatrix layers
\cite{chirikov,lichtenberg},
but their measure  drops exponentially
for $E<E_{c_1}$ and $E>E_{c_2}$. 

\begin{figure}
\begin{center}
\includegraphics[width=0.24\textwidth]{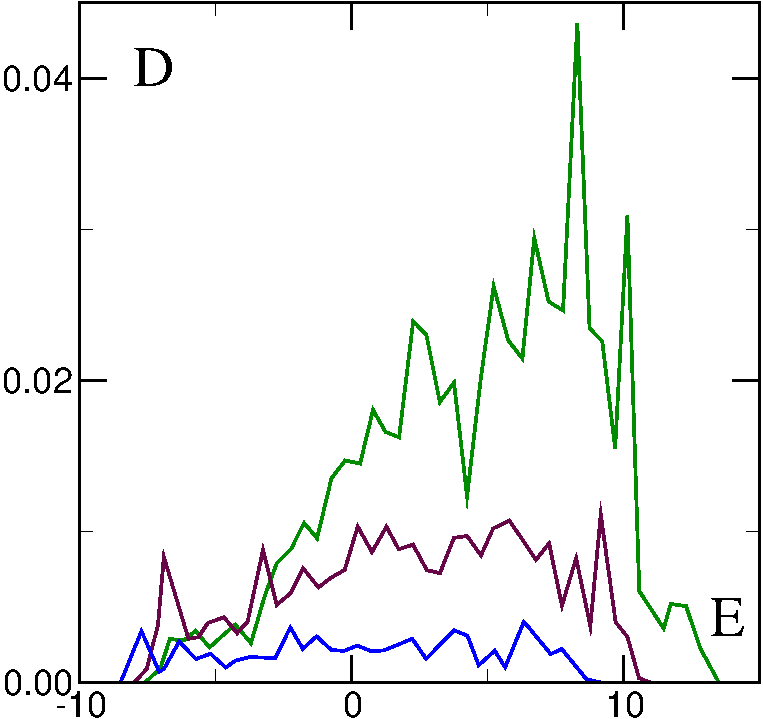}
\includegraphics[width=0.24\textwidth]{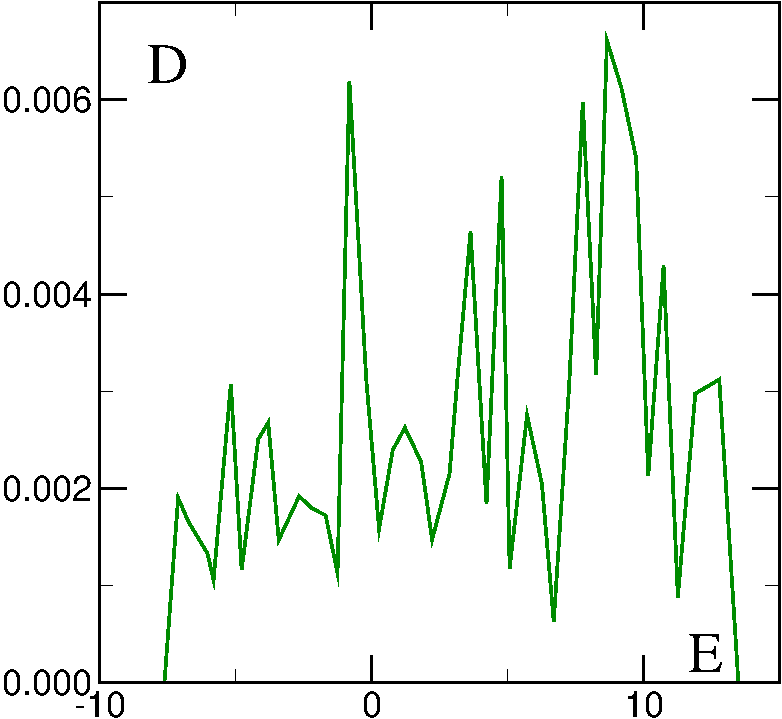}
\caption{Dependence of diffusion rate $D$ on TIP energy $E$
for same trajectories as in Fig.~\ref{fig9}
at same parameters and colors with
$\lambda=2.5$ (left panel),
$\lambda=4.5$ (right panel).
}
\label{fig10}
\end{center}
\end{figure}

The dependence of the TIP diffusion coefficient $D$ on energy
$E$, determined from the relation
$\sigma = Dt + const$, is shown in Figure~\ref{fig10}
for parameters of Figure~\ref{fig9}. 
The typical values of $D$ in 2D are
by factor $10$ smaller than those in 1D 
(see Figure~\ref{fig7}, e.g. at $\lambda=2.5, U=6$). 
We attribute this to a larger volume of chaotic motion
so that an effective pressure of one particle on another,
which allows to overcome the potential barrier,
becomes smaller and TIP needs more time to find
a narrow channel leading to a transition from one cell to another.
For $\lambda=4.5$ the diffusion $D$ 
drops approximate in $10$ times
comparing to $\lambda=2.5$ in the agreement with 
the fact that here very specific combinations of 
all coordinates are required to overcome the potential barrier.

\begin{figure}
\begin{center}
\includegraphics[width=0.24\textwidth]{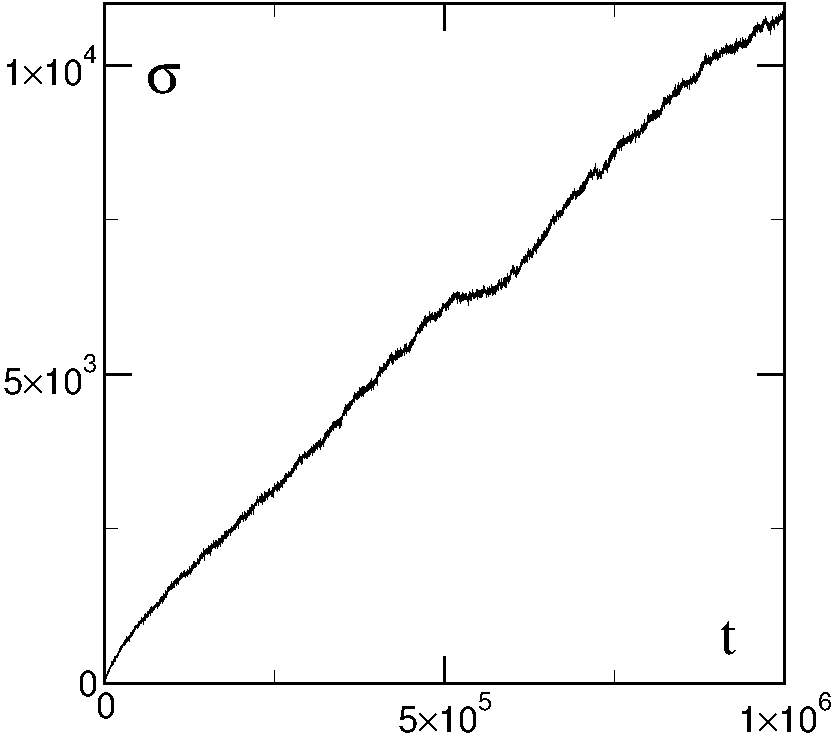}
\includegraphics[width=0.24\textwidth]{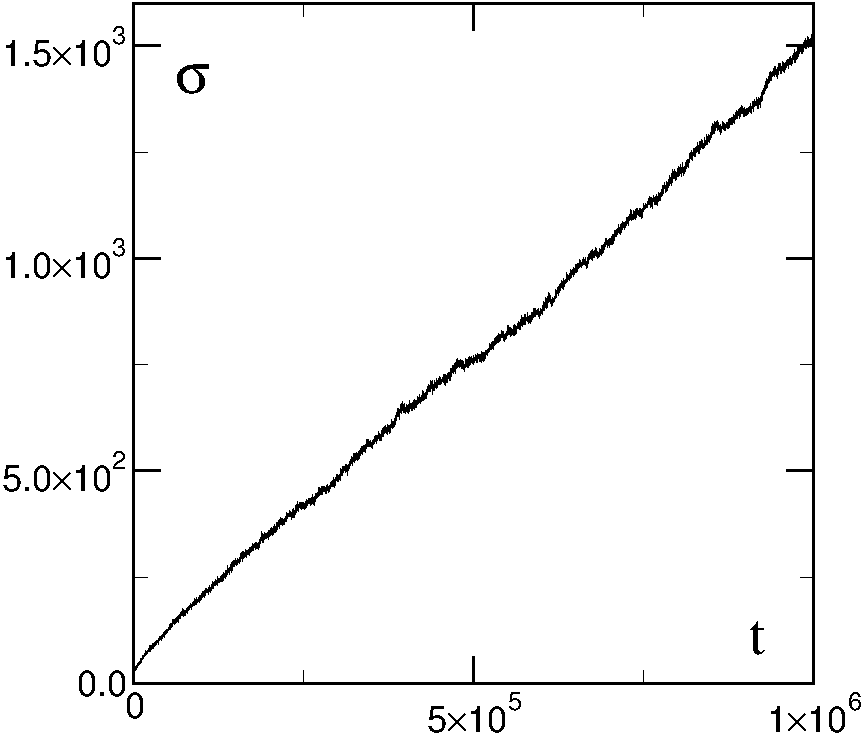}
\caption{Time dependence of the second moment
$\sigma(t)= <((x_1)^2+(x_2)^2)/2 >$
averaged over $500$ orbits taken
in a close vicinity to each other 
at $\lambda=2.5$, $U=6$, $E \approx -0.856$ (left panel)
and  $\lambda=4.5$, $U=12$, $E \approx 0.129$ (right panel).
The fit gives the diffusion rate $\sigma = Dt + const$
with $D=0.0106 \pm 3.7 \times 10^{-6}$  (left panel) and
$D=0.00143 \pm 3.8 \times 10^{-7}$ (right panel).
}
\label{fig11}
\end{center}
\end{figure}

Due to a small values of the diffusion rate $D$ the fluctuations of 
$\mu$ and $D$ are larger for 2D case comparing to 1D.
An example of a more exact computation of $D$ is shown in Figure~\ref{fig11}
where the second moment $\sigma$ is characterized by
a linear growth with time reaching rather high values.
Thus with large times and large number of trajectories the diffusion
coefficient $D$ is determined with a high precision. 
Even if $\sigma$ in Figure~\ref{fig11} has large values
corresponding to TIP displacement on a typical distance
of 16 cells the maximal distance between 
two particles remains small  $|\Delta x|/2\pi, |\Delta y|/2\pi < 2.5$
for $\lambda=2.5$ and respectively $1.5$ for $\lambda=4.5$
at $t=4 \times 10^5$.

\begin{figure}
\begin{center}
\includegraphics[width=0.24\textwidth]{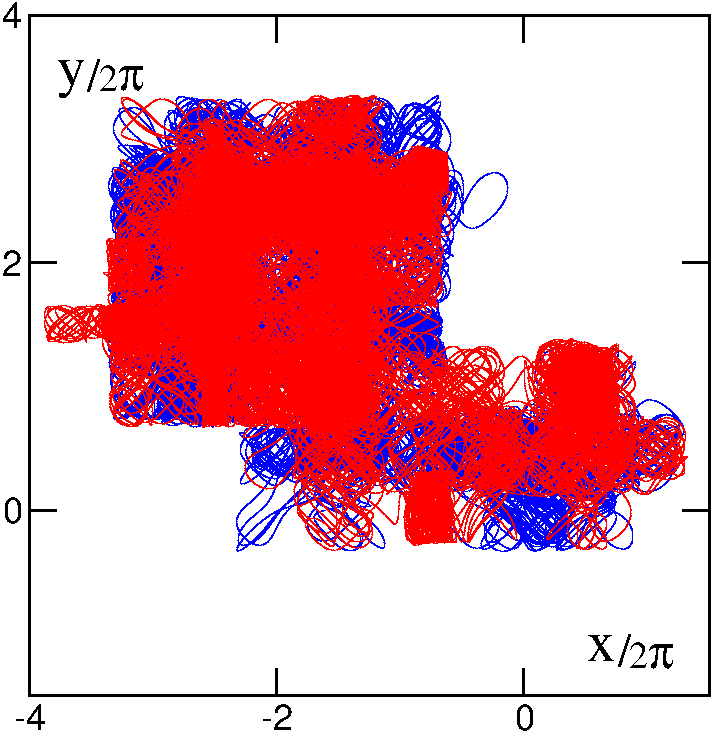}
\includegraphics[width=0.24\textwidth]{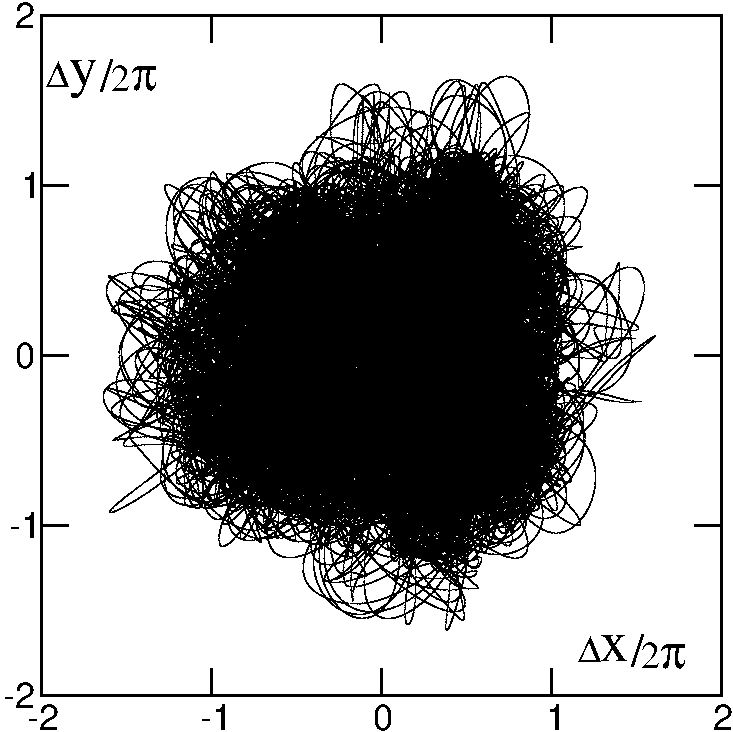}
\caption{Example of dynamics of one trajectory 
from Fig.~\ref{fig11} (left panel at $\lambda=2.5, U=6, E=-0.856$) 
for time $t \leq 10^4$ is shown on $(x,y)$ plane
with first particle in blue and second particle in red colors 
(left panel); the evolution of distance between particles
$\Delta x= x_2-x_1$, $\Delta y = y_2 - y_1$
is shown on right panel.
}
\label{fig12}
\end{center}
\end{figure}

An example of complex chaotic motion of TIP on small
scales at $t \leq 10^4$ and $\lambda=2.5, U=6$ is shown in
Figure~\ref{fig12}. During this time particles make
a displacement of up to 5 cells while the distance between them
remains less than 2 cells.

\begin{figure}
\begin{center}
\includegraphics[width=0.24\textwidth]{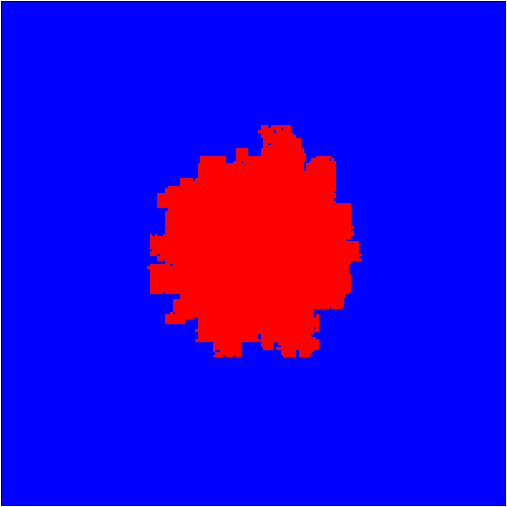}
\includegraphics[width=0.24\textwidth]{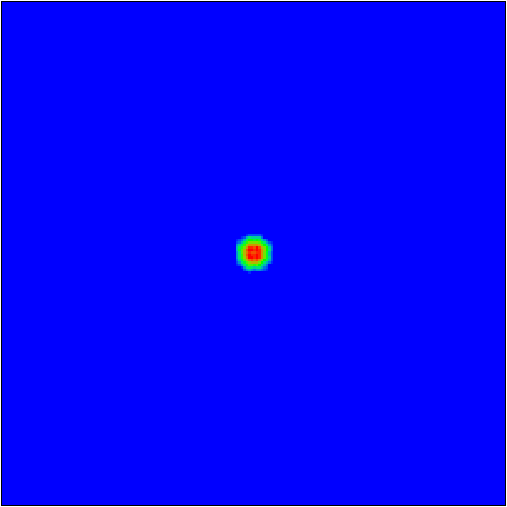}\\
\includegraphics[width=0.24\textwidth]{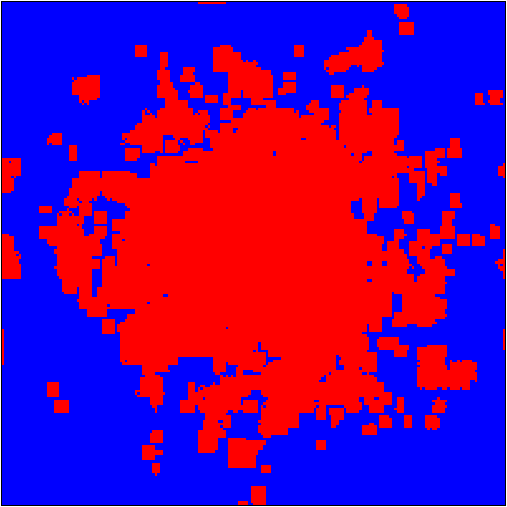}
\includegraphics[width=0.24\textwidth]{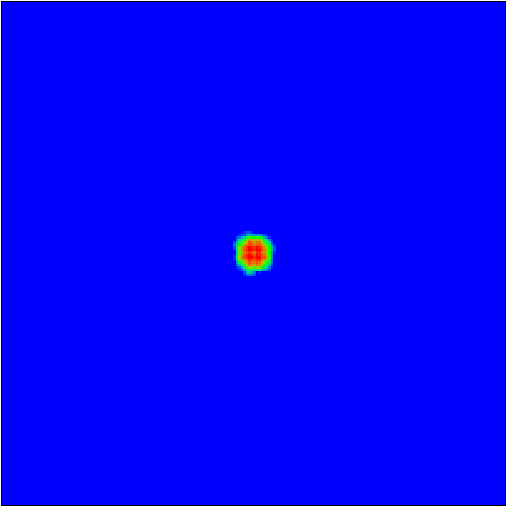}
\caption{Density distribution
of  TIP at parameters  $\lambda=2.5, U=6, E=-0.856$
averaged over $10^3$ trajectories.
Left column: density of charge (sum over each particle)
in the plane $(x/2\pi, y/2\pi)$ 
at time $t=10^4$ at top panel
and $t=10^5$ at bottom panel
(average over time interval $\delta t=10^3$); 
right column: density distribution
over distance between particles
$\Delta x/2\pi = (x_2-x_1)/2\pi,
\Delta y/2\pi = (y_2-y_1)/2\pi$
at time $t=10^4$ at top panel
and  $t=10^5$ at bottom panel
(average over all times from zero to $t$).
The panels show the squares 
$-16 \leq x/2\pi, y/2\pi \leq 16$ (left),
$-16 \leq \Delta x/2\pi, \Delta y/2\pi \leq 16$
(right);
at $t=0$ TIP are located in one cell;
color changes from blue (zero) to red (maximum).
}
\label{fig13}
\end{center}
\end{figure}

The spreading of $10^3$ TIP trajectories with time
is illustrated in Figure~\ref{fig13}.
While with time TIP cover larger and larger area
in $(x,y)$ plane their relative distance in 
number of cells remains less than 2.5.
Isolated fragments visible in 
the bottom left panel of Figure~\ref{fig13}
are generated by trajectories which gained
a large TIP displacement but then
due to fluctuations 
are stacked in a few cells
on the time interval of averaging $\delta t= 10^3$.

A more detailed verification of the validity
of the Fokker-Planck equation in 2D
(similar to Figure~\ref{fig5})
requires averaging over larger number of 
trajectories and larger times due to
smaller values of the diffusion coefficient $D$ in 2D
and we do not perform such a comparison here
considering that the linear growth of the second moment
on large times in Figure~\ref{fig11}
provides a sufficient confirmation of
the diffusive TIP propagation in the 2D Harper model.

Finally, we note that the chaotic dynamics
is also found in numerical simulations with the
kinetic TIP spectrum ${p_1}^2/2 + {p_2}^2/2$,
instead of $2\cos p_1 + 2\cos p_2$ in (\ref{eq4})
or (\ref{eq7}). However, for such a spectrum 
at energies above the potential barrier it is
found that particles become separated 
from each other with one escaping to infinity
and another one remaining trapped by a potential
so that a joint propagation of TIP
is not detected in the cases which 
have been studied numerically.

\section{Discussion}
\label{sec5}

The presented results clearly show that 
for the classical Harper model in 1D and 2D
the interactions between two particles
lead to emergence of chaos and chaotic 
propulsion of TIP with their
diffusive spreading over the whole lattice.
Such an interaction induced diffusion 
appears in the regime when without interactions
particle motion is bounded to one cell of periodic potential.
For diffusive delocalized TIP dynamics the
relative distance between particles
remains always smaller than the 
size of one to three periodic
cells.  In this sense the delocalized chaotic
TIP represent a well defined new type of quasiparticles
which we call {\it chaons} due to the chaos origin of their
diffusive propagation over the whole system.
 
The chaon diffusion takes place 
inside a certain  energy 
range of delocalized chaotic dynamics
$E_{c_1} < E < E_{c_2}$ for a moderate
interaction strength being comparable or even smaller
then the one-particle energy band.
Our results show that about a half of all energy 
band width can belong to delocalized dynamics.
The energies $E_{c_1}, E_{c_2}$ play a role of mobility edge
in energy.
Of course, at very weak interactions 
the KAM integrability is restored 
keeping the TIP dynamics bounded inside one 
periodic cell. Since the Harper model
is strongly nonlinear the chaotic delocalization
takes plays for a broad range of interactions
including the long range Coulomb interaction and
a short range interaction.

The question about the quantum manifestations of 
this classical chaon delocalization
is open for further investigations.
It is possible that the quasiballistic FIKS
pairs in the 1D quantum Harper model \cite{flach,fiks1d}
appear as a result of quantization of 
classical chaons. Indeed, it is known that
in the quasiperiodic quantum systems,
which are chaotic and diffusive in the
classical limit, the quantization can
create quasiballistic delocalized states
like it is the case in the kicked Harper
model (see results and Refs. in
\cite{lima,ketzmerick,prosen,artuso}).
However, the investigations of quantum chaons 
and their possible relation with FIKS pairs
requires further studies.
The results presented in  \cite{flach,fiks1d}
are mainly done for the Hubbard interaction
or a short range interaction on a 
discrete lattice and a semiclassical 
limit for such interactions is not so straightforward.
Indeed,  the semiclassical regime requires
a smooth potential variation and 
a large number of quantum states $n_q$ inside one potential 
period. It is possible that the semiclassical
description can work even at moderate
values of $n_q \sim 3$ but a detailed analysis of
semiclassical description of such cases
is required.  The quantum 
interference effects may lead to the quantum localization
of chaon diffusion in a similar way
as for the Anderson localization of TIP in
1D and 2D (see discussions for TIP in disordered 
potential in 
\cite{dlstip,imry,pichard,dlscoulomb}).
However, in the care of quasiperiodic potential,
appearing for irrational $\hbar$ values,
the situation is rather nontrivial
as show the results of subdiffusive spreading
for TIP in the 2D quantum Harper model \cite{fiks2d}.

The experimental investigations of 
diffusive chaons in the Harper model
look to be promising. Indeed, the Aubry-Andr\'e
transition has been observed already with
cold atoms in optical lattices and it has
been shown that the interaction effects play 
here an important role \cite{roati,modugno,bloch}.
The experimental progress with cold ions in optical lattices
(see e.g. \cite{blatt,vuletic1,vuletic2})
makes possible to study delocalized chaons 
with Coulomb interactions.

The author thanks  Klaus Frahm for fruitful discussions
of TIP properties and  Vitaly Alperovich for discussions
of possible realizations of considered Hamiltonian in semiconductor
heterostructures. 



\begin{thebibliography}{99}
\bibitem{harper} P.G.~Harper, Proc. Phys. Soc. London Sect. A {\bf 68}, 874 \& 879 (1955).
\bibitem{azbel} M.Y.~Azbel, Sov. Phys. JETP {\bf 19}, 634 (1964).
\bibitem{hofstadter} D.R.~Hofstadter, Phys. Rev. B {\bf 14}, 2239 ͑(1976).
\bibitem{aubry} S.~Aubry and G.~Andr\'e, Ann. Israel Phys. Soc.
                {\bf 3}, 133 (1980).
\bibitem{sokoloff} J.B.~Sokoloff, Phys. Rep. {\bf 126}, 189 (1985).
\bibitem{lana1} S.Y.~Jitomirskaya, Ann. Math. {\bf 150}, 1159 (1999).
\bibitem{lana2} J.~Bourgain, and S.~Jitomirskaya,
        Invent. Math. {\bf 148}, 453 (2002).
\bibitem{lana3} A.Avila, S.~Jitomirskaya and C.A.Marx, Ann. Math. 
         arXiv:1602.05111v1[math-ph] (2016).
\bibitem{geisel} T.~Geisel, R.~Ketzmerick, and G.~Petschel,
                Phys. Rev. Lett. {\bf 66}, 1651 (1991).
\bibitem{austin} M.~Wilkinson, and E.J.~Austin,
                Phys. Rev. B {\bf 50}, 1420 (1994).
\bibitem{dlsharper} D.L.~Shepelyansky, Phys. Rev. B {\bf 54}, 14896 (1996).
\bibitem{barelli} A.~Barelli, J.~Bellissard, Ph.~Jacquod, and D.L.~Shepelyansky,
                 Phys. Rev. Lett. {\bf 77}, 4752 (1996).
\bibitem{orso} G.~Dufour, and G.~Orso, 
              Phys. Rev. Lett. {\bf 109}, 155306 (2012).
\bibitem{dlstip}  D.L.~Shepelyansky, Phys. Rev. Lett. {\bf 73}, 2607 (1994).
\bibitem{imry} Y.~Imry, Europhys. Lett. {\bf 30}, 405 (1995).
\bibitem{pichard} D.~Weinmann, A. Muller--Groeling, 
                 J.-L.~Pichard, and K.~Frahm,
                  Phys. Rev. Lett. {\bf 75}, 1598 (1995).
\bibitem{frahm1995} K. Frahm, A. Muller--Groeling, J.-L. Pichard, and
	D.~Weinmann, Europhys. Lett. {\bf 31}, 169 (1995).
\bibitem{vonoppen} F. von Oppen, T.~Wetting, and J.~Muller,
                Phys. Rev. Lett. {\bf 76}, 491 (1996).
\bibitem{borgonovi} F.~Borgonovi, and D.L.~Shepelyansky,
               J. de Physique I France {\bf 6}, 287 (1996).
\bibitem{frahm1999}  K.M.~Frahm, Eur. Phys. J. B, {\bf 10}, 371 (1999).
\bibitem{dlscoulomb} D.L.~Shepelyansky, Phys. Rev. B {\bf 61},  4588 (2000).
\bibitem{lagesring} J.~Lages, and D.L.~Shepelyansky, 
                 Eur. Phys. J. B {\bf 21}, 129 (2001).
\bibitem{frahm2016} K.M.~Frahm, Eur. Phys. J. B, {\bf XX}, XXX (2016);
                arXiv:1602.08257[cond-mat.quant-gas] (2016).
\bibitem{flach} S.~Flach, M.~Ivanchenko, and R.~Khomeriki,
               Europhys. Lett. {\bf 98}, 66002 (2012).
\bibitem{fiks1d} K.M.~Frahm, and D.L.~Shepelyansky,
               Eur. Phys. J. B {\bf 88}, 337 (2015).
\bibitem{fiks2d} K.M.~Frahm, and D.L.~Shepelyansky,
               Eur. Phys. J. B {\bf 89}, 8 (2016).
\bibitem{roati} G.~Roati, C.~D`Errico, L.~Fallani, M.~Fattori, 
            C.~Fort, M.~Zaccanti,
                G.~Modugno, M.~Modugno, and M.~Inguscio,
                Nature {\bf 453}, 895 (2008).
\bibitem{modugno} E.~Lucioni, B.~Deissler, L.~Tanzi, G.~Roati,
                  M.~Zaccanti, M.~Modugno, M.~Larcher, F.~Dalfovo, M.~Inguscio,
                  and G.~Modugno,
                 Phys. Rev. Lett. {\bf 106}, 230403 (2011).
\bibitem{bloch} M.~Schreiber, S.S.~Hodgman, P.~Bordia, 
                  H.~Luschen,
                 M.H.~Fischer, R.~Vosk, E.~Altman,
                 U.~Schneider, and I.~Bloch,
                 Science {\bf 349}, 842 (2015).
\bibitem{bloch2d} P.~Bordia, H.K.~Luschen, S.S.~Hodgman, 
                M.~Schreiber, I.~Bloch, and U.~Schneider,
                http://arxiv.org/abs/1509.00478 (2015).
\bibitem{pokrovsky} V.L.~Pokrovsky and A.L.~Talapov,
                   {\it Theory of Incommensurate Crystals},
                  Harwood, London v.1 (1984).
\bibitem{gruner} G.~Gruner, {\it Density Waves in Solids},
        Addison-Wesley Publ. Company, New York (1994).
\bibitem{brazovski} S.A.Brazovskii, {\it Ferroelectricity and 
        charge ordering in quasi-1D organic conductors}, in 
        {\it The Physics of Organic Superconductors and Conductors} 
        A.Lebed (Ed.),
        p.313, Springer-Verlag, Berlin (2008).
\bibitem{alper} V.L.~Alperovich, N.T.~Moshegov, V.A.~Tkachenko, O.A.~Tkachenko,
         A.I.Toropov, and A.S.Yaroshevich,
         Pis'ma ZhETF {\bf 70}, 112 (1999) [in Russian].
\bibitem{chirikov} B.V.~Chirikov, Phys. Reports {\bf52}, 263 (1979).
\bibitem{lichtenberg} A.J.Lichtenberg, M.A.Lieberman, 
        {\it Regular and chaotic dynamics}, Springer, Berlin (1992).
\bibitem{lima} R.~Lima, and D.L.~Shepelyansky,
        Phys. Rev. Lett. {\bf 67},  1377 (1991).
\bibitem{ketzmerick} R.~Ketzmerick, K.~Kruse, and T.~Geisel,
        Physica D {\bf 131}, 247 (1999).
\bibitem{prosen} T.~Prosen, I.~I.~Satija, and N.~Shah,
        Phys. Rev. Lett. {\bf 87}, 066601 (2001).
\bibitem{artuso} R.~Artuso, Scholarpedia {\bf 6(10)}, 10462 (2011).
\bibitem{blatt} P.~Jurcevic, P.~Hauke, C.~Maier, C.~Hempel, B.P.~Lanyon, R.~Blatt, and C.F.~Roos,
        Phys. Rev. Lett. {\bf 115}, 100501 (2015).
\bibitem{vuletic1} A.~Bylinskii, D.~Gangloff, and V.~Vuletic,
         Science {\bf 348}, 1115 (2015).
\bibitem{vuletic2} A.~Bylinskii, D.~Gangloff, I.~Counts and V.~Vuletic,
         Nature Materials doi:10.1038/nmat4601 (2016).


\end{thebibliography}
\end{document}